\documentclass[aps,prb,superscriptaddress,twocolumn,reprint]{revtex4-1}
\usepackage{graphicx,bm,times}
\usepackage{amsmath}
\usepackage{amsfonts}
\usepackage{amssymb}
\usepackage{color}
\usepackage{hyperref}
\hypersetup{
	colorlinks = true,
	allcolors = {blue}
}
\usepackage[utf8]{inputenc}
\usepackage[T1]{fontenc}
\usepackage{graphicx}
\usepackage{amsmath,amssymb}
\usepackage{lineno}
\begin{document}
	
\title{Band hybridisation at the semimetal-semiconductor transition of Ta$_2$NiSe$_5$\\enabled by mirror-symmetry breaking}
	
\author{Matthew D.\ Watson}
\email{matthew.watson@diamond.ac.uk}
\affiliation {SUPA, School of Physics and Astronomy, University of St Andrews, St Andrews KY16 9SS, United Kingdom}

\author{Igor Markovi{\'c}}
\affiliation {SUPA, School of Physics and Astronomy, University of St Andrews, St Andrews KY16 9SS, United Kingdom}
\affiliation{Max Planck Institute for Chemical Physics of Solids, N\"{o}thnitzer Strasse 40, 01187 Dresden, Germany}

\author{Edgar Abarca Morales}
\affiliation {SUPA, School of Physics and Astronomy, University of St Andrews, St Andrews KY16 9SS, United Kingdom}
\affiliation{Max Planck Institute for Chemical Physics of Solids, N\"{o}thnitzer Strasse 40, 01187 Dresden, Germany}

\author{Patrick Le F{\`e}vre}
\affiliation {Synchrotron SOLEIL, L'Orme des Merisiers, Saint-Aubin-BP48, 91192 Gif-sur-Yvette, France}

\author{Michael Merz}
\affiliation {Institute for Quantum Materials and Technologies, Karlsruhe Institute of Technology, 76021 Karlsruhe, Germany}

\author{Amir A.\ Haghighirad}
\affiliation {Institute for Quantum Materials and Technologies, Karlsruhe Institute of Technology, 76021 Karlsruhe, Germany}
	
\author{Philip D.\ C.\ King}
\email{philip.king@st-andrews.ac.uk}
\affiliation {SUPA, School of Physics and Astronomy, University of St Andrews, St Andrews KY16 9SS, United Kingdom}
                                                                                                                                                                                                                                                                                                                                                                                                                                                                                                                                             
\date{\today}

\begin{abstract}
 We present a combined study from angle-resolved photoemission and density-functional theory calculations of the temperature-dependent electronic structure in the excitonic insulator candidate Ta$_2$NiSe$_5$. Our experimental measurements unambiguously establish the normal state as a semimetal with a significant band overlap of $>$100~meV. Our temperature-dependent measurements indicate how these low-energy states hybridise when cooling through the well-known 327~K phase transition in this system. From our calculations and polarisation-dependent photoemission measurements, we demonstrate the importance of a loss of mirror symmetry in enabling the band hybridisation, driven by a shear-like structural distortion which reduces the crystal symmetry from orthorhombic to monoclinic. Our results thus point to the key role of the lattice distortion in enabling the phase transition of Ta$_2$NiSe$_5$.
	\end{abstract}
\maketitle

A key idea in condensed matter physics is how many-body interactions can lead to the emergence of novel ground states, such as Mott insulators, conventional and unconventional superconductors and density waves. The long-proposed excitonic insulator (EI) phase is a particularly intriguing example. It is an instability of a narrow-gap semiconductor or semimetal driven by the spontaneous formation and/or condensation of electron-hole bound states (i.e. excitons)~\cite{Jerome1967}. In its purest form, it is exclusively an electronically-driven many-body effect, caused by the attractive Coulomb interaction between electrons and holes, which is poorly screened due to the small carrier densities \cite{Bronold2006PRB}. However, whether the EI state occurs naturally in the solid state has proved highly controversial. Partly, this is because, in real materials, the presence of discrete crystal symmetries and finite electron-lattice coupling means that there is inevitably a structural component to any EI-like phase transition \cite{Zenker2014PRB}. For indirect band gap systems, any coupling between the valence and conduction bands will thus also generate a charge density wave (CDW). However, CDW instabilities by themselves can open up similar single-particle energy gaps without any excitonic contribution, which has led to significant controversy over the interpretation of results on putative excitonic transitions in indirect band gap semiconductors such as TiSe$_2$ \cite{Kidd2002,van_Wezel_2010,rohwer_collapse_2011,Monney2011PRL,Calandra2011,Rossnagel2011Review,Kogar1314,Watson2019PRL}.

It is therefore attractive to consider EI candidates with a small direct band gap/overlap. Ta$_2$NiSe$_5$ is currently the most prominent material in this category, and a number of its physical properties are consistent with an EI-like transition \cite{Lu2017NComms,Mor2017PRL,Okazaki2018NComms,Lee2019PRB,Kaneko2013PRB,Seki2014PRB}. Transport and optical spectroscopy indicate a near-zero band gap at high temperatures but the opening of a band gap at low temperatures \cite{Lu2017NComms,Larkin2017PRB}, with a clear resistivity anomaly at $T_c$=327~K \cite{DiSalvo1986JLCM}. Previous low-temperature angle-resolved photoemission (ARPES) measurements show an anomalous flattened dispersion of the valence band maximum, indicating that it is a hybridised ground state \cite{Mor2017PRL,Fukutani2019PRL}, while a number of time-resolved studies have found evidence for a fast timescale associated with electronic interactions \cite{Okazaki2018NComms,Mor2017PRL}. Nevertheless, there is also a non-negligible coupling to the lattice. The 327~K second-order phase transition is accompanied by a lowering of the space group symmetry, as a shear-like distortion reduces the symmetry from orthorhombic to monoclinic \cite{Sunshine1985,Nakano2017IUCrJ,Nakono2018PRB}. Fano-like lineshapes in optical spectroscopy suggested a strong exciton-phonon coupling \cite{Larkin2017PRB}, while time-resolved studies also show coherent oscillations associated with a coupling to phonon modes which persist for a longer time than the fast electronic timescales \cite{Werdehausen2018SciAdv}. Thus, despite the absence of a CDW, it is again important to gauge the relative importance of the structural and excitonic contributions. 

In this letter, we show that the structural component of the phase transition indeed plays a crucial role. Using polarisation-dependent ARPES, we show that the valence and conduction bands derive from states with opposite mirror parity in the orthorhombic phase. This prohibits any band hybridisation at high temperature, but in the monoclinic phase, two mirror symmetries are broken, allowing hybridisation between the Ni-derived valence bands and Ta-derived conduction bands. Our \textit{ab-initio} calculations yield a hybridised semiconducting ground state in the monoclinic phase with a band gap of $\sim$0.12 eV, similar to that observed experimentally. Our results thus show how band and structural symmetries are key to understanding the unusual instability of Ta$_2$NiSe$_5$.  

\begin{figure*}
	\centering
	\includegraphics[width=0.8\linewidth]{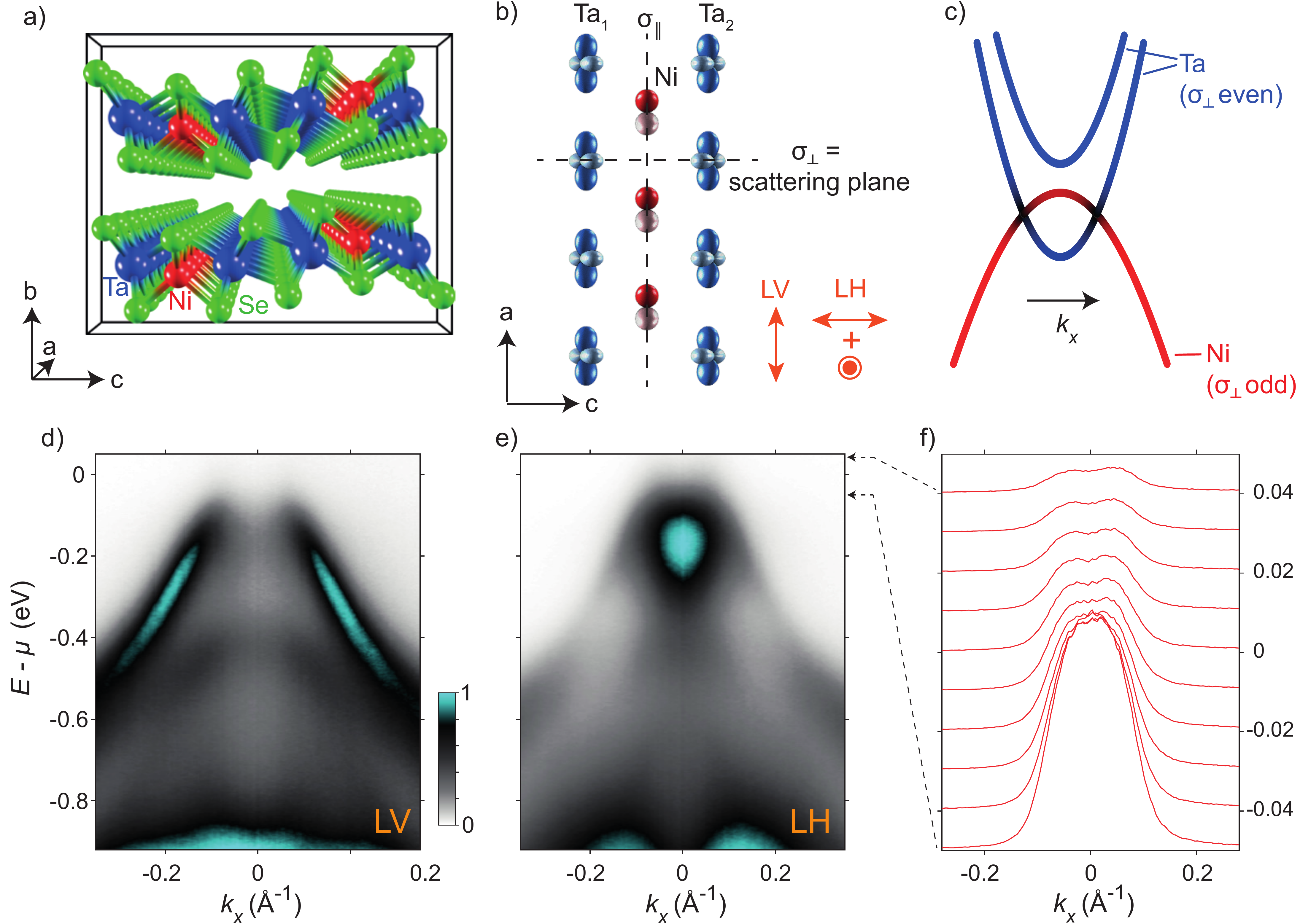}
	\caption{a) Crystal structure of Ta$_2$NiSe$_5$. b) Subset of Ta $5d$ and Ni $3d$ orbital states which are relevant to the low-energy bands structure, following Ref.~\cite{Kaneko2013PRB}. Note their opposite parities with respect to the mirror plane perpendicular to the chains (horizontal dashed line). c) Schematic of expected low-energy band structure in the orthorhombic phase. d,e) ARPES measurements taken at $T$=350~K (above $T_c$), measured along the $\Gamma$-X direction using (d) LV and (e) LH polarised light. Note that at such elevated temperatures, the chemical potential may be significantly different from the $T$=0 charge neutrality point where the valence and conduction bands overlap. The photoemission geometry is represented in (b), where the crystallographic chains ($a$ axis) are parallel to the analyser entrance slit, while the scattering plane is coincident with the $\sigma_{\perp}$ mirror plane. f) Momentum distribution curves extracted from the region shown in (e), demonstrating an electron-like band dispersing upwards away from the intense ``blob" of spectral weight at the conduction band minimum.} 
	\label{fig:fig1}
\end{figure*}

High-quality single crystals were synthesised by the chemical vapour transport method. ARPES data were measured at the CASSIOP{\'E}E beamline at the Soleil synchrotron. Samples were cleaved {\it in situ} and measurements were performed at temperatures between 113~K and 350~K. The data presented in the main text were all measured with a photon energy of 24 eV, probing close to a bulk $\Gamma$ plane along $k_y$ (see Fig.~SM1~in the Supplemental Material, SM \cite{SM}), and with linear horizontal or vertical polarised light, respectively. Density functional theory (DFT) calculations were performed within the Wien2k package \cite{wien2k}, using the modified Becke-Johnson (mBJ) exchange-correlation potential \cite{Koller2012_mBJ}, and including spin-orbit coupling. The Brillouin zone was sampled by a 35$\times$35$\times$7 $k$-mesh.  The muffin-tin radius $R_{\text{MT}}$ for all atoms was chosen such that the product with the maximum modulus of reciprocal vectors $K_{\text{max}}$ becomes $R_{\text{MT}}K_{\text{max}}$ = 7.0.  Experimental crystal structures were used as input. For the orthorhombic phase, we used the crystal structure determined at 400~K~\cite{Nakano2017IUCrJ}, while for the low-temperature monoclinic phase we used the recent structural refinement obtained at 30~K by Nakano~{\it et al.} \cite{Nakano2018PRB}. 

The crystal structure of Ta$_2$NiSe$_5$ consists of parallel chains of Ta and Ni atoms, with Se ions providing the Ta and Ni with octahedral and tetrahedral coordination, respectively, as shown in Fig.~\ref{fig:fig1}(a) and detailed in Table SM-1 in the SM \cite{SM}. A quasi-1D electronic structure can thus be expected, although the relatively low symmetry implies a large number of bands. Kaneko \textit{et al.} \cite{Kaneko2013PRB,KanekoThesis} identified a limited subset of orbital states that dominate the low-energy band structure, depicted schematically in Fig.~\ref{fig:fig1}(b,c), with two conduction bands derived from Ta $5d$ orbitals, and the uppermost valence band originating from $3d$ orbitals from the Ni chains (neglecting the contributions from Se atoms for simplicity). 

\begin{figure}
	\centering
	\includegraphics[width=\linewidth]{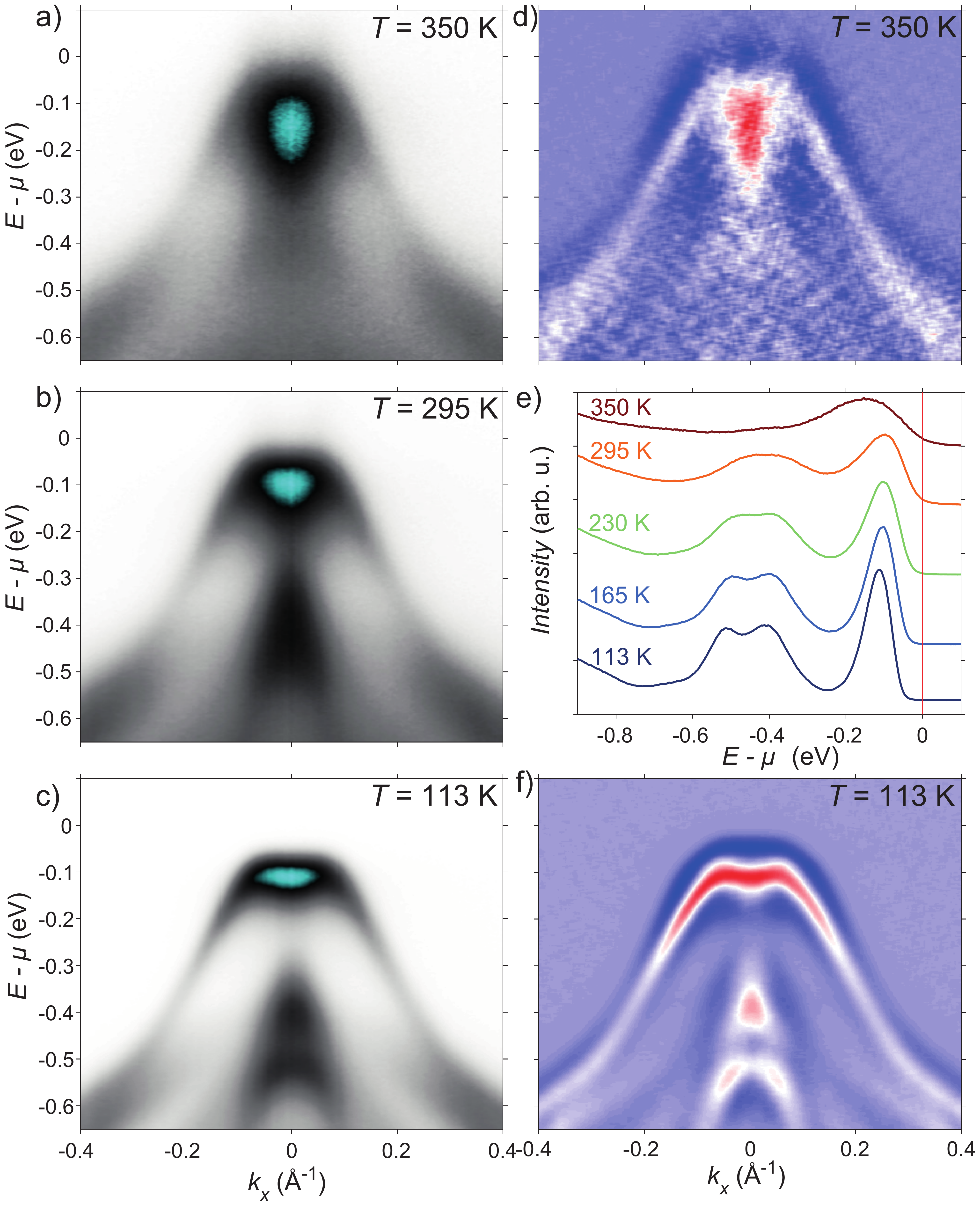}
	\caption{(a-c) Temperature-dependent ARPES data, measured along the $\Gamma$-X direction for temperatures of (a) 350~K, (b) 295~K, and (c) 113~K. (d,f) ``Curvature" analysis of panels (a) and (c), respectively. (e) Energy distribution curves (EDCs) of the data at $k_x$ = 0. All data obtained at $h\nu{}$~=~24~eV, in LH polarization.}
	\label{fig:fig2}
\end{figure}

The two mirror symmetries marked in Fig.~\ref{fig:fig1}(a) (parallel and perpendicular to the chain direction) are especially important. In our photoemission measurements, the $\sigma_{\perp}$ mirror plane is coincident with the scattering plane of our measurement geometry, meaning that it sets the photoemission selection rules. Following well-established symmetry analysis~\cite{Damascelli2003}, ARPES measurements performed with linear vertical (LV) and horizontal (LH) incident beam polarisations (see Fig.~\ref{fig:fig1}(b) for the experimental geometry) will selectively probe electronic states with odd and even parity with respect to $\sigma_{\perp}$ \footnote{Strictly, in our geometry, the selection rule holds only along the $\Gamma-Z$ line, i.e. at $k_x$=0, but deviations from this should not be large for the momentum range considered here.}. From this, we can identify that the valence band in Fig.~1(d) has odd parity with respect to $\sigma_\perp$. In contrast, for measurements using LH polarisation in Fig.~\ref{fig:fig1}(e), the spectrum is dominated by the bottom of the conduction band at $\Gamma$, indicating that the conduction band has even parity relative to the $\sigma_{\perp}$ scattering plane.  

Two conduction bands would be expected since there are two Ta chains for each Ni chain (Fig.~\ref{fig:fig1}(c)). As some weak hopping processes are possible between the Ta chains, they split into bonding and antibonding states with opposite parity with respect to the mirror plane parallel to the chain, $\sigma_{\parallel}$. Our measurements demonstrate that the lower-energy state, most likely the $\sigma_{\parallel}$-odd state, has its band minimum $\approx\!175$~meV below the chemical potential, forming an electron-like state which disperses upwards to cross the chemical potential as evident in momentum distribution curves (MDCs, Fig.~\ref{fig:fig1}(f)). Crucially, due to the different mirror parity of the valence and conduction bands with respect to $\sigma_\perp$, they cannot hybridise at $\Gamma$. No band gap can form, and the normal state of Ta$_2$NiSe$_5$ is a robust semimetal.

Our experimental ARPES data indicate a number of qualitative differences between the spectra measured just above and below $T_c$ (Fig.~\ref{fig:fig2}(a-c), further highlighted with ``curvature" analysis~\cite{curvature} in Fig.~\ref{fig:fig2}(d,f)). First, below $T_c$ the electron-like dispersion crossing $E_F$ disappears.  Second, at low temperatures we no longer observe any bands overlapping and there is only a single band dispersion within 0.3 eV of $\mu$ in the occupied states. That state exhibits a characteristic ``M" shaped dispersion \cite{Wakisaka2009PRL,Wakisaka2012JSNM,Seki2014PRB,Mor2017PRL,Fukutani2019PRL}, which becomes sharper as the temperature is further lowered (Fig.~\ref{fig:fig2}(c,f)). In the shallow local minimum at the $\Gamma$ point the band has a positive effective mass, and remains intense in our measurements using LH-polarised light, suggesting that here it largely retains the Ta atomic character of the high-temperature conduction band for small $k_x$. However, as it disperses further away from $\Gamma$ there is a crossover towards the dispersion corresponding to the original Ni-derived valence band. 

\begin{figure*}
	\centering
	\includegraphics[width=0.85\linewidth]{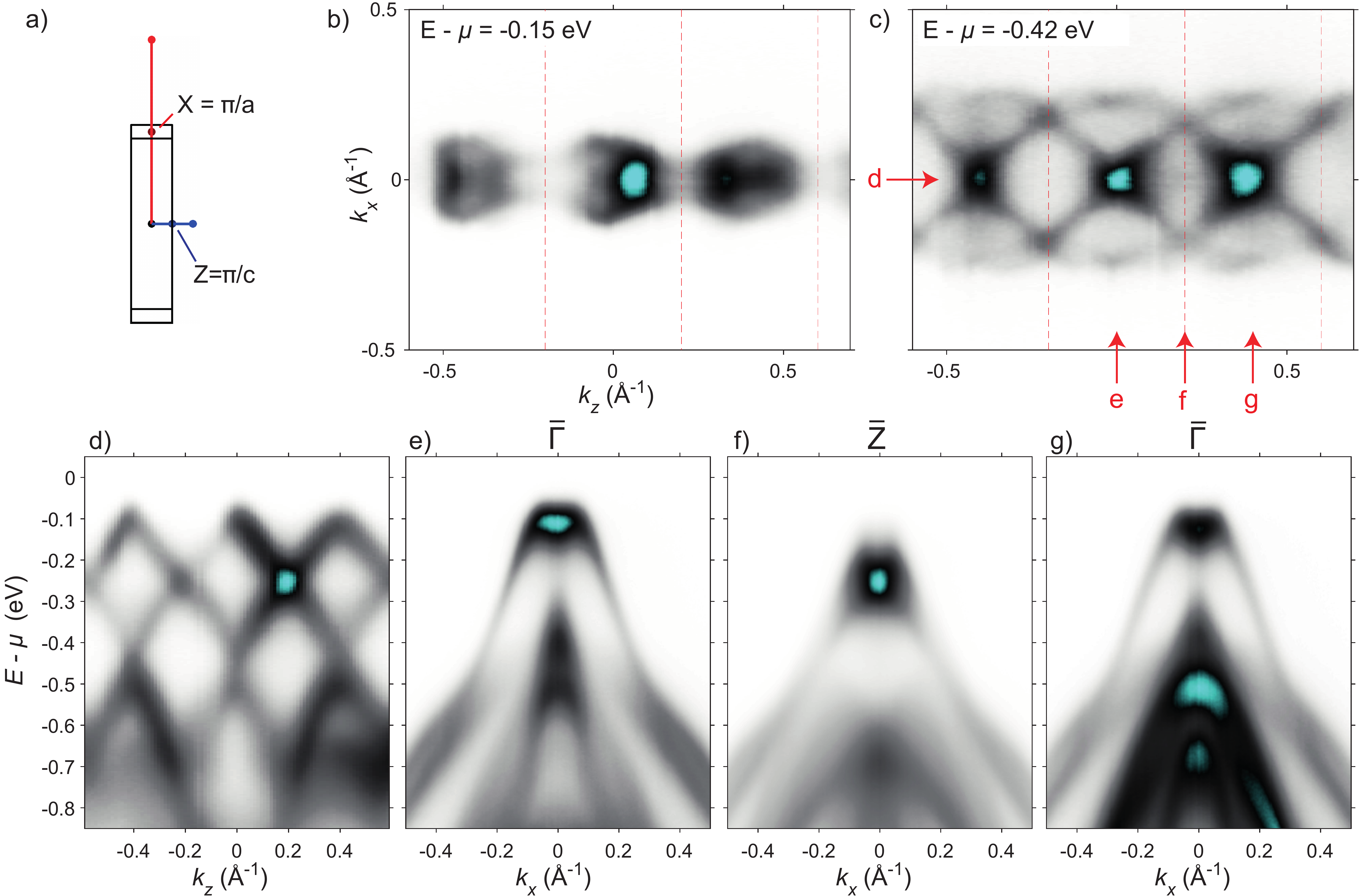}
	\caption{a) Projection of the 3D Brillouin zone in the $k_z,k_x$ plane; the high-symmetry points are labeled (see also Fig.~SM-1 in SM). b) Intensity map obtained for a constant energies of -0.15 eV, a little below the flattened hybridised state found at $\Gamma$, showing closed 2D contours. c)  Equivalent measurement at -0.42 eV below $\mu$, also clearly showing features dispersing in both $k_z$ and $k_x$ directions. Dashed red lines are the Brillouin zone boundaries. d) Dispersion in the $k_z$ direction, obtained from the mapping data, and (e-g) High-symmetry dispersions as indicated in (c). All data measured at $h\nu{}$ = 24 eV in LH polarization.}
	\label{fig:fig3}
\end{figure*}

Thus, our data imply that from immediately below $T_c$, a hybridisation process gaps all states at the chemical potential,  $\mu$, mediating a transition from a semimetallic state at high temperature to a semiconducting state below $T_c$, i.e. a semimetal-semiconductor transition. This leaves a particularly strongly hybridised state as the uppermost valence band. By 113~K the valence band maximum is pushed down to 0.11 eV below $\mu$. Since only the occupied states are probed by ARPES, we can determine only a lower limit of the band gap. We estimate this to be 0.13 eV, otherwise hints of the conduction band would have been visible at the elevated temperatures here. These observations are consistent with the optical gap of 0.16 eV  \cite{Lu2017NComms,Larkin2017PRB}, and with analysis of scanning tunneling spectra which show a similar gap size, with the chemical potential pinned closer to the conduction band minimum than the valence band maximum~\cite{Lee2019PRB}. 

So far, we have considered only the band structure in the vicinity of the $\Gamma$ point, and considered only the dispersions along the chain directions. However, there are possible hopping processes between the Ta-Ni-Ta tri-chains; as a result, the bands also show some dispersion in the perpendicular directions, both in-plane and in the out-of-plane direction. In Fig.~\ref{fig:fig3}(b,c) we present constant energy maps in the $k_z-k_x$ plane, for the monoclinic phase. For a perfectly one-dimensional electronic structure, the dispersions would be independent of $k_z$, but the maps clearly show features dispersing in two dimensions. In Fig.~\ref{fig:fig3}(b), at an energy slightly below the flattened state, the intensity is localised around the $\Gamma$ point, demonstrating a substantial 2D character of the uppermost valence band. At a lower energy in Fig.~\ref{fig:fig3}(c) we similarly find rather 2D features. The dispersion along $k_z$ (perpendicular to the usual measurement in the $k_x$ direction) in Fig.~\ref{fig:fig3}(d) shows periodic dispersions covering several Brillouin zones, with the dispersions showing with typical bandwidths of $\sim$0.2 eV.
	
At the Z point, shown in Fig.~\ref{fig:fig3}(f) the uppermost valence band is found at a significantly higher binding energy than at the $\Gamma$ point (Fig.~\ref{fig:fig3}(e,g) and Fig.~\ref{fig:fig2}). However, the bands here do not show the flattened, hybridized band structure that is characteristic of the $\Gamma$ point. The larger spectral gap in Fig.~\ref{fig:fig3}(f) should therefore not be interpreted as a larger hybridization effect at Z than at $\Gamma$; rather, the uppermost valence bands here are simply at a lower energy due to hopping processes perpendicular to the chains, that are largely unrelated to the monoclinic distortion. A similar dispersion is observed in the S-analogue, Ta$_2$NiS$_5$ \cite{Mu2018}. Thus, for considering the effects of the band hybridisation at the 327~K phase transition, we focus on the bands around the $\Gamma$ point, where the  bands overlap in the tetragonal phase and thus dominant hybridisation effects can be expected, as are indeed found in the monoclinic phase. However, these measurements do emphasize that the one-dimensional picture often used for Ta$_2$NiSe$_5$ is not fully representative of the whole electronic structure. The other direction, the out-of-plane $\Gamma-Y$ direction, corresponds to $k_\perp$ in our ARPES geometry, and can be probed by varying the photon energy; we present photon-energy-dependent data in the SM Fig.~\ref{fig:fig1} \cite{SM}, but the variation are relatively limited, at least for the low-energy Ta and Ni states.

\begin{figure*}
	\centering
	\includegraphics[width=\linewidth]{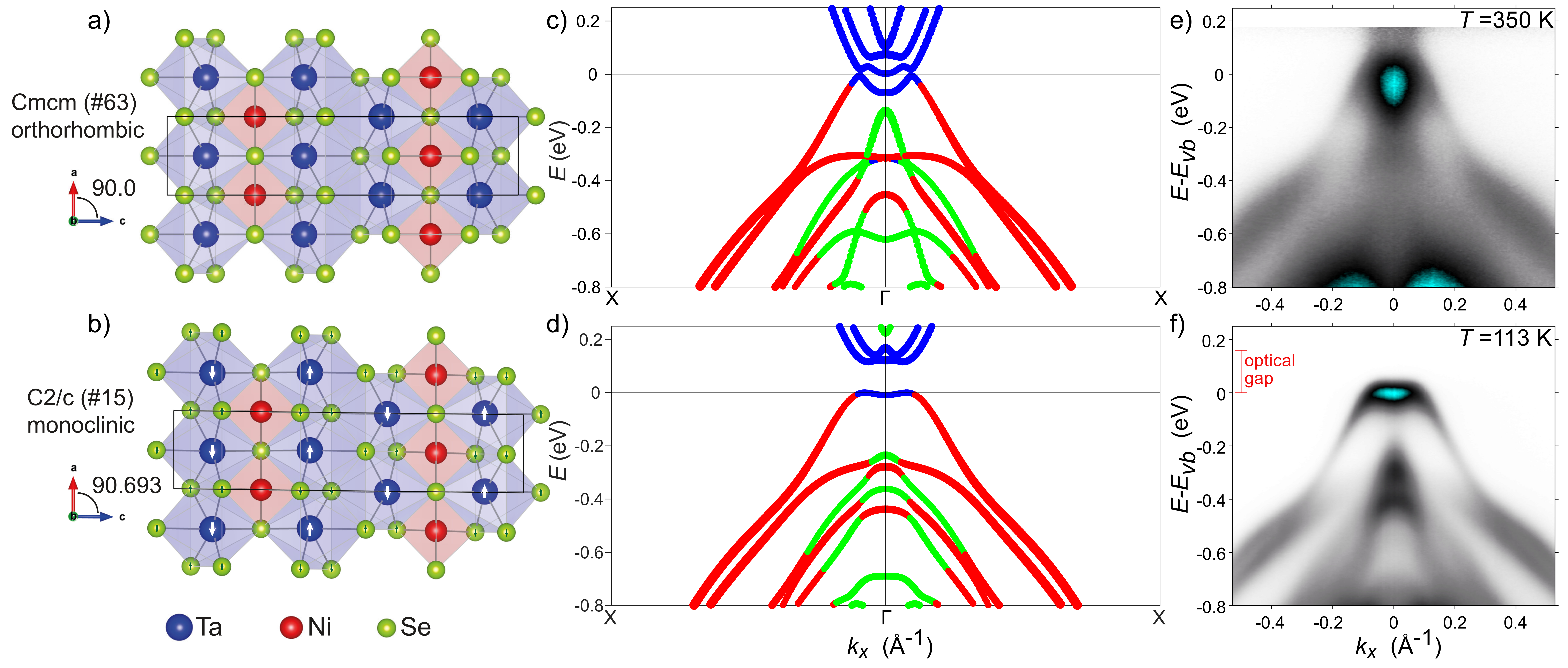}
	\caption{(a,b) Top view of the crystal structures in the orthorhombic and monoclinic phases. The white and green arrows in (b) represent the atomic displacements arising due to the shear distortion. This results in a loss of the $\sigma_\perp$ and $\sigma_\parallel$ mirror symmetries for the low-temperature structure. (c,d) Corresponding DFT calculations using the mBJ exchange potential in the orthorhombic and monoclinic phases. Blue, red and green colours are used to represent where the maximum atomic character of the band is Ta,Ni, or Se, respectively. (e,f) ARPES data in the orthorhombic and monoclinic phases, reproduced from Fig,~\ref{fig:fig2} but now referenced to the valence band maximum. In (e), the data are divided by the Fermi function to better highlight the spectral weight above $E_F$. The optical gap of 0.16 eV \cite{Lu2017NComms} is shown in (f) for comparison.}
	\label{fig:fig4}
\end{figure*}

We now turn to the nature of the order parameter driving the transition. From a symmetry perspective, in order to develop a band gap by hybridising the valence and conduction states, the $\sigma_\perp$ mirror symmetry present in the parent orthorhombic phase must be broken. This is satisfied by the shear-like displacements of the Ta atoms in the monoclinic and semiconducting phase below $T_c$ (see arrows in Fig.~\ref{fig:fig4}(b)), providing an intuitive understanding of how the structural changes are intimately related to the low-energy electronic structure. Nevertheless, an excitonic order could in principle also break these discrete symmetries, and the lattice deformation could be a secondary effect \cite{Kaneko2013PRB}, or the two orders may be comparable in magnitude and cooperative. Thus, symmetry analysis alone cannot determine the nature of the order parameters, and one must carefully consider the energy scales of the problem.

It has previously been assumed that the magnitude of the band hybridisation expected from the structural distortion terms was small, thus favouring the interpretation first proposed by Kaneko \textit{et al} \cite{Kaneko2013PRB} in which the leading role is played by a condensation of excitons. In a similar spirit, Lee \textit{et al} \cite{Lee2019PRB}, performed DFT calculations in the monoclinic phase and found only a very small gap of 0.03 eV, too small to explain the experimental observations, and thus apparently implying significant many-body interactions beyond DFT. However, for our DFT calculations, presented in Fig.~\ref{fig:fig4}(c,d), we find a substantial band gap of 0.12 eV for the low-temperature crystal structure, coming rather close to the experimental value. The main difference is the choice of functional; we adopt the modified Becke-Johnson exchange potential \cite{Koller2012_mBJ} which is known to lead to improved estimates of band gaps in semiconductors, and similarly band overlaps in semimetals.  The success of this functional here is confirmed by the good agreement between our experimental data in the orthorhombic phase and the calculations for the high-temperature crystal structure, which correctly reproduce a realistic semimetallic state (Fig.~\ref{fig:fig4}(c,e)).  Importantly, for the low-temperature structure in the monoclinic phase, the DFT calculation reproduces the hybridised valence band rather well, showing a shallow M-shaped dispersion and an evolution from Ta character around $k_x$=0 to Ni character further away from $\Gamma$ (Fig.~\ref{fig:fig4}(d,f)), with a magnitude of the corresponding hybridisation gap that is comparable to that observed experimentally. It is also notable that only one of the two conduction band branches becomes hybridised in the monoclinic phase, reflecting the opposite parity of the bonding and anti-bonding conduction band states with respect to $\sigma_\parallel$. This results in the presence of both a strongly-hybridised and a non-bonding conduction band, as also evident in the model of Ref.~\cite{Kaneko2013PRB}.

As well as the hybridisation of the Ta and Ni-derived conduction and valence bands, our DFT calculations highlight that the displacements of Se atoms are also relevant, in particular pushing the light, second highest valence band with dominant Se character down in energy. A similar shift is evident in our experimental temperature-dependent measurements (Fig.~\ref{fig:fig2}), where the state at around 0.4~eV below the chemical potential becomes sharper and shifts to higher energy with decreasing temperature. This reflects the fact that both Se and Ta are displaced below $T_c$ (Fig.~\ref{fig:fig4}(b))~\cite{Nakano2018PRB,Yan2019}: the shear displacements of the Ta atoms is most relevant for the low energy dispersions, but there is additionally an antiparallel motion of the Se atoms which coordinate the Ta sites, allowing for further hybridisations, and providing an additional energetic incentive for the phase transition.

We therefore conclude that DFT calculations taking into account the monoclinic distortions can yield a semiconducting ground state with band dispersions closely resembling the ARPES data and a band gap close to the experimental value. While this does not preclude that electron-hole correlations may still contribute to driving the phase transition in Ta$_2$NiSe$_5$, our findings highlight that it is the symmetry-breaking structural distortions that can ultimately explain the energetics of the phase transition here. These allow new hopping pathways and hybridise the valence and conduction bands, opening up an energy gap and giving a sufficiently large free energy gain to enable a second order phase transition above room temperature.

\textit{Note added: During the final stages of preparing our manuscript, a theoretical preprint appeared which makes similar statements regarding the symmetry aspects of the monoclinic distortion to those made here \cite{Mazza2019_arxiv}. }

\section{Acknowledgements}
We thank Andreas Rost for useful discussions, and Craig Topping for assisting with sample characterisation. We gratefully acknowledge support from The Leverhulme Trust and The Royal Society. IM and EAM acknowledge studentship support through the International Max-Planck Research School for the Chemistry and Physics of Quantum Materials. The contribution from M.M. was supported by the Karlsruhe Nano Micro Facility (KNMF). We thank SOLEIL synchrotron for access to the CASSIOP{\'E}E beamline which contributed to the results presented here.

%

\end{document}